\documentclass{PoS}

\newcommand{\be}{\begin{equation}}
\newcommand{\ee}{\end{equation}}
\newcommand{\bea}{\begin{eqnarray*}}
\newcommand{\eea}{\end{eqnarray*}}

\usepackage{epsfig}

\PoS{PoS(LAT2005)269}

\title{Dynamical supersymmetry breaking and phase diagram of the lattice Wess-Zumino model }

\ShortTitle{Dynamical supersymmetry breaking and phase diagram of the lattice Wess-Zumino model }

\author{Matteo Beccaria and Gian Fabrizio De Angelis\\
        INFN, Sezione di Lecce, and
Dipartimento di Fisica dell'Universit\`a di Lecce,
Via Arnesano, ex Collegio Fiorini, I-73100 Lecce, Italy\\
        E-mail: \email{beccaria@le.infn.it} , \email{deangelis@le.infn.it} }

\author{Massimo Campostrini\thanks{IFUP-TH 2005/23}\\
        INFN, Sezione di Pisa, and
Dipartimento di Fisica ``Enrico Fermi'' dell'Universit\`a di Pisa,
Via Buonarroti 2, I-56125 Pisa, Italy \\
        E-mail: \email{campostrini@df.unipi.it}}

\author{\speaker{Alessandra Feo}\thanks{UPRF-2005-04}\\
        Dipartimento di Fisica, Universit\`a di Parma and INFN Gruppo Collegato di Parma, Parco
  Area delle Scienze 7/A, 43100 Parma, Italy \\
        E-mail: \email{feo@fis.unipr.it}}

\abstract{We study dynamical supersymmetry breaking and the transition point 
by non-perturbative lattice
techniques in a class of two-dimensional $N=1$ Wess-Zumino model. 
The method is based on the calculation of rigorous lower 
bounds on the ground state 
energy density in the infinite-lattice limit.  
Such bounds are useful in the discussion of supersymmetry phase 
transition. The transition
point is determined with this method and then compared with recent results based on 
large-scale Green Function Monte Carlo simulations with
good agreement.  }

\FullConference{XXIIIrd International Symposium on Lattice Field Theory\\
                  25-30 July 2005\\
		 Trinity College, Dublin, Ireland}

\begin{document}

\section{Introduction}
An important issue in the study of supersymmetric models is the occurrence of 
non-perturbative dynamical supersymmetry breaking~\footnote{See \cite{feo} for 
recent reviews and a complete list of references.}.
The problem can be studied in the $N=1$ Wess-Zumino 
model that does not involve gauge fields and is thus a simple theoretical laboratory.
Since the  breaking of supersymmetry is closely related to the symmetry properties of the ground state, 
we will adopt a Hamiltonian formulation of the model.

Let us remind the (continuum) $N=1$ super algebra, 
$\{Q_\alpha,Q_\beta\} = 2(\not{P} C)_{\alpha\beta}$. 
Since $P_i$ are not conserved on the lattice, the super algebra 
is explicitly broken by the lattice discretization.
A very important advantage of the Hamiltonian formulation is the
possibility of conserving exactly a subset of the full super algebra \cite{elitzur}.
Specializing to $1+1$ dimensions, in a Majorana basis 
$\gamma_0 = C = \sigma_2$, $\gamma_1 = i \sigma_3$, the algebra becomes: 
$Q_1^2 = Q_2^2 = P^0 \equiv H$ and $\{Q_1,Q_2\} = 2 P^1 \equiv 2 P$.
On the lattice, since $H$ is conserved but $P$ is not, we can pick up 
one of the supercharges, say, 
$Q_1$, build a discretized version $Q_L$ and define the lattice Hamiltonian 
to be $H = Q_L^2$.
Notice that $Q_1^2 = H$ is enough to guarantee $E_0 \ge 0$. 
The explicit lattice model is built by considering a spatial 
lattice with $L$ sites. On each site we place 
a real scalar field $\varphi_n$ together with its conjugate momentum $p_n$ such 
that $[p_n, \varphi_m] = -i\delta_{n,m}$.  The associated fermion is a Majorana fermion
$\psi_{a, n}$ with $a=1,2$ and 
$\{\psi_{a, n}, \psi_{b, m}\} = \delta_{a,b}\delta_{n,m}$, 
$\psi_{a,n}^\dagger = \psi_{a,n}$. 

The continuum 2-dimensional Wess-Zumino model is defined by the supersymmetric 
generators involving the superpotential $V(\varphi)$,
\be
Q_{1,2} = \int  dx \left[ p(x) \psi_{1,2}(x)
    - \Bigg( {\partial\varphi\over\partial x} \pm V(\varphi(x)) \Bigg) \psi_{2,1}(x) \right],
\ee
where $\varphi(x)$ is a real scalar field and $\psi(x)$ is a Majorana fermion.
The discretized supercharge is~\cite{elitzur,ranft}
\be
Q_L = \sum_{n=1}^L\left [
p_n\psi_{1,n}-\left(\frac{\varphi_{n+1}-\varphi_{n-1}}{2}
+ V(\varphi_n) \right)\psi_{2,n}\right] 
\ee
and the Hamiltonian takes the form 
\be
H = Q_L^2 = {1\over2} \sum_{n=1}^L \Bigg[ \pi_n^2 + 
\left({\phi_{n+1} - \phi_{n-1}\over2} + V(\phi_n)\right)^{\!2}
 - (\chi^\dagger_n\chi_{n+1} + h.c.) + (-1)^n V'(\phi_n)
\left(2\chi^\dagger_n\chi_n-1\right) \Bigg] 
\ee
where we replace the two Majorana fermion operators with a single Dirac operator
$\chi$ satisfying canonical anticommutation rules.

The problem of predicting the pattern of supersymmetry breaking is 
not easy. In principle, the form of $V(\varphi(x))$ is enough to 
determine whether supersymmetry is broken or not.
At least at tree level supersymmetry is broken if and only if $V$ has no zeros.
The Witten index \cite{witten} can help in the analysis: If $V(\varphi)$ has an odd 
number of zeroes then $I \not= 0$ and supersymmetry is unbroken. If $V(\varphi)$ 
has an even number of zeroes, when $I=0$ we can not conclude anything.  
An alternative non-perturbative analysis for the case $I=0$ is thus welcome. The 
simplest way to analyze the pattern of supersymmetry breaking for a given $V$ 
is to compute the ground state energy $E_0$ through numerical simulations 
and/or strong coupling expansion. 
On the lattice, accurate numerical results are available~\cite{wz1,beccaria}, although
a clean determination of the supersymmetry breaking transition remains rather elusive.
All the predictions for the model with cubic prepotential, $V=\varphi^3$, 
indicated unbroken supersymmetry. 
Dynamical supersymmetry breaking in the model with quadratic prepotential 
$V=\lambda_2\varphi^2 + \lambda_0 $ was studied performing
numerical simulations \cite{wz1} along a line of constant $\lambda_2$, confirming the 
existence of two phases:
a phase of broken supersymmetry with unbroken discrete $Z_2$ at high $\lambda_0$ 
and a phase of unbroken supersymmetry with broken $Z_2$ at low $\lambda_0$, 
separated by a single phase transition.

On the other hand, from the strong coupling analysis what comes out is the following: 
for odd $q$, strong coupling and weak coupling expansion results agree 
and supersymmetry is expected to be unbroken \cite{wz1}.
This conclusion gains further support from the non vanishing value 
of the Witten index \cite{witten}.
For even $q$ in strong coupling, the ground state has a positive energy 
density also for $L \to \infty $ and supersymmetry appears to be broken. 
In particular, for $V=\lambda_2\varphi^2 + \lambda_0$, weak coupling 
predicts unbroken supersymmetry for $\lambda_0 < 0$, whereas strong coupling 
prediction gives broken supersymmetry for all $\lambda_0$.

\section{Numerical Simulations and Discussion}
We used two different approaches to investigate the pattern of dynamical supersymmetry 
breaking.
In the first one, \cite{wz1,beccaria}, the numerical simulations were performed using the 
Green Function Monte Carlo (GFMC) algorithm and strong coupling expansion.
The GFMC is a method that computes a numerical representation of the ground state wave function 
on a finite lattice with $L$ sites in terms of the states carried by an ensemble of $K$ walkers.
Numerical results using the GFMC algorithm for the odd prepotential confirm unbroken 
supersymmetry.

A more interesting case is the even prepotential. 
When $V= \lambda_2 \varphi^2 + \lambda_0$ and for fixed 
$\lambda_2= 0.5$, 
we may expect (in the $L\to\infty$ limit) a phase transition at
$\lambda_0=\lambda_0^{(c)}(\lambda_2)$ 
separating a phase of broken supersymmetry and unbroken $Z_2$
(high $\lambda_0$) from a phase of unbroken supersymmetry  
and broken $Z_2$ (low $\lambda_0$).  

\begin{figure}
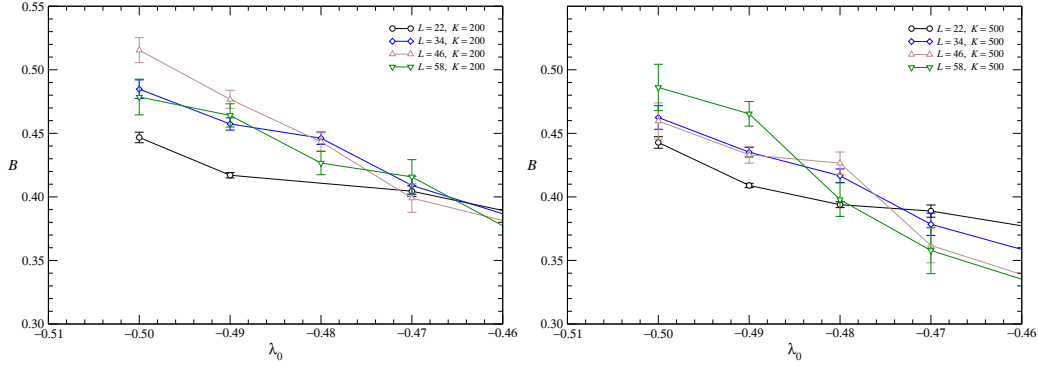

\epsfig{file=B.K_200.eps,width=0.45\textwidth,angle=0}
\epsfig{file=B.K_500.eps,width=0.45\textwidth,angle=0}
\caption{The Binder cumulant $B$ vs. $\lambda_0$ for $K=200$ and $K=500$. Here $\lambda_0^{(c)}=-0.48\pm0.01$.}
\label{2}
\end{figure}

The usual technique for the study of a phase transition is the
crossing method applied to the Binder cumulant, $B$.
The crossing method consists in plotting $B$ vs.\ $\lambda_0$ for
several values of $L$. 
The crossing point $\lambda_0^{\rm cr}(L_1,L_2)$, determined 
by the condition
$ B(\lambda_0^{\rm cr},L_1) = B(\lambda_0^{\rm cr},L_2)$
is an estimator of $\lambda_0^{(c)}$. 
The value obtained is showed in Fig. \ref{2} and corresponds to 
$\lambda_0^{(c)}=-0.48\pm0.01$ \cite{wz1}.
The main source of systematic errors in this method is the need to extrapolate to infinity
both $K$ and $L$. For this reason, an independent method to test the numerical results of 
\cite{wz1} is welcome. 

The second method is based on the calculation of rigorous lower bounds 
on the ground state energy density in the {\em infinite-lattice\/} 
limit \cite{wz2,proc}. Such bounds are useful in the discussion of 
supersymmetry breaking as follows:  
The lattice version of the Wess-Zumino model conserves
enough supersymmetry to prove that the ground state has a non 
negative energy density $\rho\ge 0$, as its continuum limit.
Moreover the ground state is supersymmetric if and only 
if $\rho=0$, whereas it breaks (dynamically) supersymmetry 
if $\rho>0$. Therefore, if an exact positive lower bound 
$\rho_{\rm LB}$ is found with $0< \rho_{\rm LB} \le \rho$,
we can claim that supersymmetry is broken.

The idea is to construct a sequence $\rho^{(L)}$ 
of exact lower bounds representing the ground state
energy densities of modified lattice Hamiltonians describing 
a cluster of $L$ sites and converging to $\rho^{(L)} \to \rho$ 
in the limit $L\to\infty$.  
The bounds $\rho^{(L)}$ can be computed numerically on a 
finite lattice with $L$ sites.
The relevant quantity for our analysis is the ground state 
energy density $\rho$ evaluated on the infinite lattice limit 
$ \rho = \lim_{L\to\infty}\frac{E_0(L)}{L} $.
It can be used to tell between the two phases of the model: 
supersymmetric with $\rho=0$ or broken with $\rho > 0$.

In Ref. \cite{wz2} we presented how to build a sequence of bounds $\rho^{(L)}$
which are the ground state energy density of the Hamiltonian H with modified 
couplings on a cluster of $L$ sites:
given a translation-invariant Hamiltonian $H$ on a regular lattice it 
is possible to obtain a lower 
bounds on its ground state energy density from a cluster decomposition 
of $H$, i.e., given a suitable finite sublattice $\Lambda$, 
it is possible to introduce a modified Hamiltonian 
$\widetilde H$ restricted to $\Lambda $ such that its energy density 
$\rho_\Lambda$ bounds $\rho$ from below. The difference between 
$H$ and $\widetilde H$ amounts to a simple rescaling
of its coupling constants. 
The only restriction on $H$ being that its interactions must have a 
finite range \cite{wz2}.

We compute numerically $\rho^{(L)}$ at various values of the 
cluster $L$: if we find $\rho^{(L)}> 0$ for some $L$ we 
conclude that we are in the broken phase.
We know that $\rho^{(L)} \to \rho$ for $L \to \infty$ and the 
study of  $\rho^{(L)}$  as a function of {\em both} $L$ and 
the coupling constants permit the identification of the phase 
in all cases. The calculation of  $\rho^{(L)}$ is numerically 
feasible because it requires to determine the ground state energy
of a Hamiltonian quite similar to $H$ and defined on a finite lattice
with $L$ sites.

\begin{figure}
\begin{center}
\epsfig{file=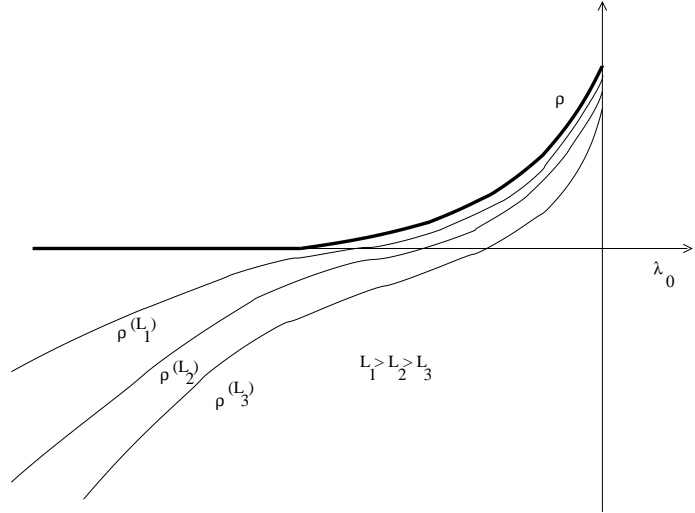,width=0.60\textwidth,angle=0}
\end{center}
\caption{Qualitative plot of the functions $\rho(\lambda_0)$ and 
$\rho^{(L)}(\lambda_0)$.}
\label{3}
\end{figure}

To test the effectiveness of the proposed bound and its relevance 
to the problem of locating the supersymmetry transition in the Wess-Zumino model
we study in detail the case of a 
quadratic prepotential $V= \lambda_2 \varphi^2 + \lambda_0$ at a fixed value 
$\lambda_2 = 0.5$ \cite{wz2}. 
An argument by Witten \cite{witten} suggest the existence of a negative number 
$\lambda_0^\ast$ such that $\rho(\lambda_0)$ is positive
when $\lambda_0 > \lambda_0^\ast$ and it vanishes for 
$\lambda_0 < \lambda_0^\ast$. 
$\lambda_0^\ast$ is the value of $\lambda_0$ in which dynamical 
supersymmetry breaking occurs. 
In Fig. \ref{3} we show a qualitative pattern of the curves 
representing $\rho^{(L)}(\lambda_0)$. We see that a single zero 
is expected in $\rho^{(L)}(\lambda_0)$
at some $\lambda_0 = \lambda_0(L)$. Since $\lim_{L\to\infty}\rho^{(L)} = \rho$, 
we expect that $\lambda_0(L)\to\lambda_0^*$ for $L\to \infty$
allowing for a determination of the critical coupling $\lambda_0^*$.
The continuum limit of the model is obtained by following a Renormalization Group
trajectory that, in particular, requires the limit $\lambda_2 \to 0$ \cite{wz1}.

\begin{figure}
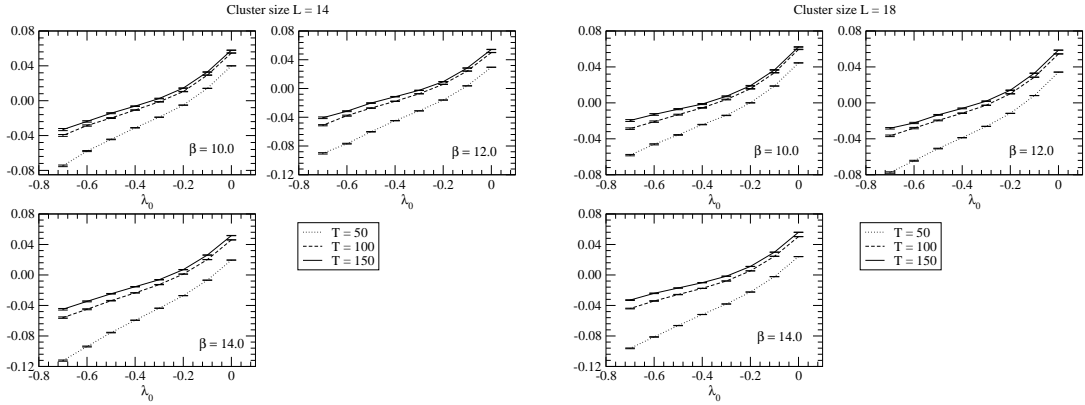

\vskip 2.3mm
\begin{center}
\epsfig{file=Plot.L14.raw.eps,width=0.45\textwidth,angle=0} \, \, \, \,
\epsfig{file=Plot.L18.raw.eps,width=0.45\textwidth,angle=0}
\end{center}
\caption{Plot of the energy lower bound $\rho^{(L)}(\beta, T)$ at $L=14$ and $L=18$.}
\label{1}
\end{figure}

\begin{figure}
\vskip 2.3mm
\begin{center}
\epsfig{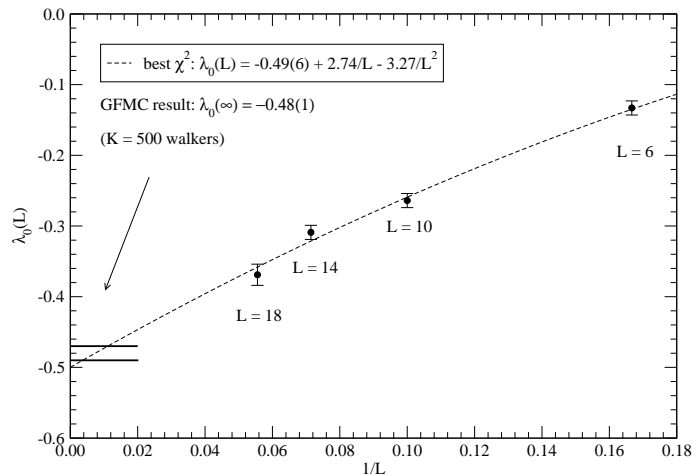}
\end{center}
\caption{Plot of $\lambda_0(L)$ vs. $1/L$ for $L=6,10,14,18$. 
The best fit with a quadratic polynomial in $1/L$ gives $\lambda_0^* = -0.49\pm 0.06$ that should 
compared with the best GFMC result obtained with $K=500$ walkers. }
\label{4}
\end{figure}

The properties of the bound $\rho^{(L)}(\lambda_0)$ guarantee that for  
$L$ large enough it must have a single zero $\lambda_0^*(L)$ converging to $\lambda_0$ as 
$L\to\infty$. In any case for each $L$ we can claim that $\lambda_0^* > \lambda_0^*(L)$.
To obtain the numerical estimate of $\rho^{(L)}(\lambda_0)$ we 
used the world line path integral (WLPI) algorithm.
The WLPI algorithm computes numerically the quantity 
$ \rho^{(L)}(\beta, T) = \frac{1}{L}\frac{\mbox{Tr}\{ H\ (e^{-\frac{\beta}{T} 
H_1}e^{-\frac{\beta}{T} H_2})^T\}}
{\mbox{Tr}\{ (e^{-\frac{\beta}{T} H_1}e^{-\frac{\beta}{T} H_2})^T\}}$
where the Hamiltonian for a cluster of $L$ sites is written as $H=H_1+H_2$, 
by separating in a convenient way the various bosonic and fermionic operators 
in the subhamiltonians $H_1$ and $H_2$ \footnote{we do not report $H_1$ and $H_2$ here,
see Ref. \cite{wz2} for details.}.
The desired lower bound is obtained by the double extrapolation
$\rho^{(L)} = \lim_{\beta\to\infty}\lim_{T\to\infty} \rho^{(L)}(\beta, T)$,
with polynomial convergence $\sim 1/T$ in $T$ and exponential in $\beta$.
Numerically, we determined 
$\rho^{(L)}(\beta, T)$ for various values of $\beta$ and $T$ 
and a set of $\lambda_0$ that should include the transition point, 
at least according to the GFMC results. In Fig. \ref{1} we plot 
the function $\rho^{(L)}(\beta, T)$ for the cluster sizes $L = 14$ and $L=18$, various $\beta$ 
and $T=50$, $100$, $150$.
Here we see that the energy lower bound behaves as 
expected: it is positive around $\lambda_0 = 0$ and 
decreases as $\lambda_0$ moves to the left. At a certain unique point $\lambda_0^*(L)$, 
the bound vanishes and 
remains negative for $\lambda_0 < \lambda_0^*(L)$. This means that supersymmetry 
breaking can be excluded 
for $\lambda_0 > \min_L\lambda_0^*(L)$. Also, consistency of the bound means that 
$\lambda_0^*(L)$ must converge
to the infinite-volume critical point as $L\to\infty$. Since the difference between 
the exact Hamiltonian and the 
one used to derive the bound is ${\cal O}(1/L)$, we can fit $\lambda_0^*(L)$ 
with a polynomial in $1/L$. This is 
shown in Fig. \ref{4} where we also show the GFMC result. 
The best fit with a parabolic function 
gives $\lambda_0^* = -0.49\pm 0.06$ \cite{wz2} quite in agreement with the previous 
$\lambda_{0, \rm GFMC}^* = -0.48\pm0.01$ \cite{wz1}.
In conclusion, both methods 
reported here are quite in agreement and confirm the existence of two phases 
separated by a single phase transition at $\lambda_0$ for the 
quadratic prepotential.

Acknowledgments: The calculations have been done on a PC Cluster at the Department
of Physics of the University of Parma.

\end{document}